\documentclass[onecolumn]{aastex63}

\usepackage{float}
\usepackage{amsmath}
\usepackage{afterpage}

\shorttitle{Science Impact MUSE}
\shortauthors{Roth}

\begin{document}

\title{Scientific Impact of novel Instrumentation: the Case of MUSE}

\correspondingauthor{Martin M. Roth}
\email{mmroth@aip.de}

\author[0000-0003-2451-739X] {Martin M. Roth}
\affiliation{Leibniz Institute for Astrophysics Potsdam (AIP), An der Sternwarte 16, 14482 Potsdam, Germany}

\begin{abstract}
In the process of transforming science cases into a viable and affordable design for a novel instrument, there is the problem 
of how to gauge their scientific impact, especially when they end up in competing top level requirements that can be incompatible with each other. This research note presents a case study for scientific impact of the integral field spectrograph MUSE in terms of number of refereed publications from 2014 to 2024 as a figure of merit, broken down by different research areas. The analysis is based on the Basic ESO Publication Statistics service (BEPS) and NASA’s Astrophysics Data System (ADS). 
\end{abstract}

  
\section{Introduction}
\label{sec:intro}
It is an almost trivial statement that novel instruments with features that open new parameter space are destined to enable new, often fundamental discoveries. Technical examples include the introduction of photoelectric photometry in 1913, and, 50 years later, CCDs \citep{Roth2023}. Examples for spectacular discoveries are the confirmation of supermassive Black Holes in the Milky Way and in M87, also Black Hole binary mergers detected with gravitational wave detectors \citep{Bambi2020}. While such achievements have confirmed predictions from theory, other examples of innovative instrumentation have led to changes of paradigms, e.g., the recent discovery with JWST that the fraction of  disk galaxies at high redshift is much larger than previously thought from the irregular galaxies seen by HST \citep{Ferreira+2023}. Yet another thread is the impact of survey facilities, e.g., the Sloan Digital Sky Survey  \citep{York+2000}.

One can ask the question whether a facility has been successful in {\em confirming the predictions for a specific experiment}, efficiently supporting a {\em diverse range of research areas}, or enabling {\em findings that were totally unexpected} at the time of the design. 

To better answer such questions, and as an attempt to review the scientific impact of an innovative instrument, this analysis focuses on the integral field spectrograph MUSE at the ESO VLT \citep{Bacon+2010} and its first decade of operation. Integral field spectroscopy (IFS) was pioneered in the late 1980's by, e.g., \citet{Vanderriest+1987,Barden+1988}. It took more than a decade of first generation instruments until the approval for the development of MUSE, that was commissioned in 2014. Thanks to the service of the BEPS\footnote{https://www.eso.org/sci/php/libraries/pubstats/} and subsequent follow-up on ADS\footnote{https://ui.adsabs.harvard.edu/}, it has been possible to track down all refereed journal papers based on MUSE data with a link to ADS. In what follows, the analysis of a total of 1054 publications will be presented and discussed.

\section{Publication Statistics of MUSE}
\label{sec:publstat}
The BEPS offers an online portal that provides quick overview over the productivity of ESO facilities using tables and plots for the number of publications per instrument. The ESO Telescope Bibliography (telbib) is a database of refereed papers resulting from observations with ESO facilities, however excluding technical papers, simulations, and publications with no immediate science content. The criteria are clearly explained on the webpage. The telbib database is accessible in portable formats. For the purpose of this study, the VLT Instruments database was downloaded as a CVS file and imported into an EXCEL spreadsheet. The table contains a total of 1058 entries, each of which represents a publication, with information on author(s), titles, year of publication, the VLT instrument used, the ADS Bibcode, and further ancillary information. All 1058 papers were screened individually and categorized in six groups in ascending order of distance: (1) Solar System, (2) Milky Way, (3) resolved stellar populations from the Milky Way to nearby Local Volume galaxies, (4) nearby galaxies out to redshift $z=0.05$, (5) intermediate redshift $0.05<z<1.0$, and (6) high redshift $z>1$ galaxies.
Roughly half of the papers could immediately be categorized on the basis of their title. The other half needed to be looked up first on ADS to check the abstract, and if this was not sufficient, by inspection of the downloaded paper itself.
As a result, 4 publications were rejected as either erroneously associated with MUSE data, or for purely technical content. With an error rate of 0.4\% this fraction can be considered negligible.

\section{Results}
\label{sec:results}
The top panel in Fig.~\ref{fig:figure1} shows the development of productivity per year for all VLT instruments, beginning in 1999. In the year of commissioning (2014), two MUSE papers were published, followed by a nearly linear increase every year. By 2020, MUSE became the most productive instrument of the entire VLT instrument suite, yielding 201 papers in 2023.

It is interesting to break down the statistics by category, as shown in the pie chart in the bottom panel of Fig.~\ref{fig:figure1}. 
With 42\%, the overwhelming fraction of papers was published about nearby galaxies, followed by 19\% and 22\% for intermediate and high redshift galaxies, respectively. Together, the latter add up to almost exactly the same fraction as of the one for nearby galaxies. With a total of 58 papers, Solar System and Milky Way have produced only a relatively small fraction, while resolved stellar populations arrive at a fraction of 11\%. 

It is further interesting to observe the development since 2014. While there is the general trend of increase over the years, which is understandable for an observing technique that is still relatively new and more challenging in terms of data reduction and analysis, categories (1) and (2) show only a hint of this trend. However, for all of the remaining categories (3) to (6) the increase is very clear, and almost monotonic. It is fair to state that extragalactic astronomy and cosmology have profited enormously from IFS with MUSE. However, given the current oversubscription of VLT-UT4, it is likely that the trend will eventually end and a saturation must occur, simply because of limited observing time.

\section{Discussion and Conclusions}
\label{sec:conclusions}
\begin{figure}[h!]
  \centering
  \includegraphics[width=1.0\textwidth]{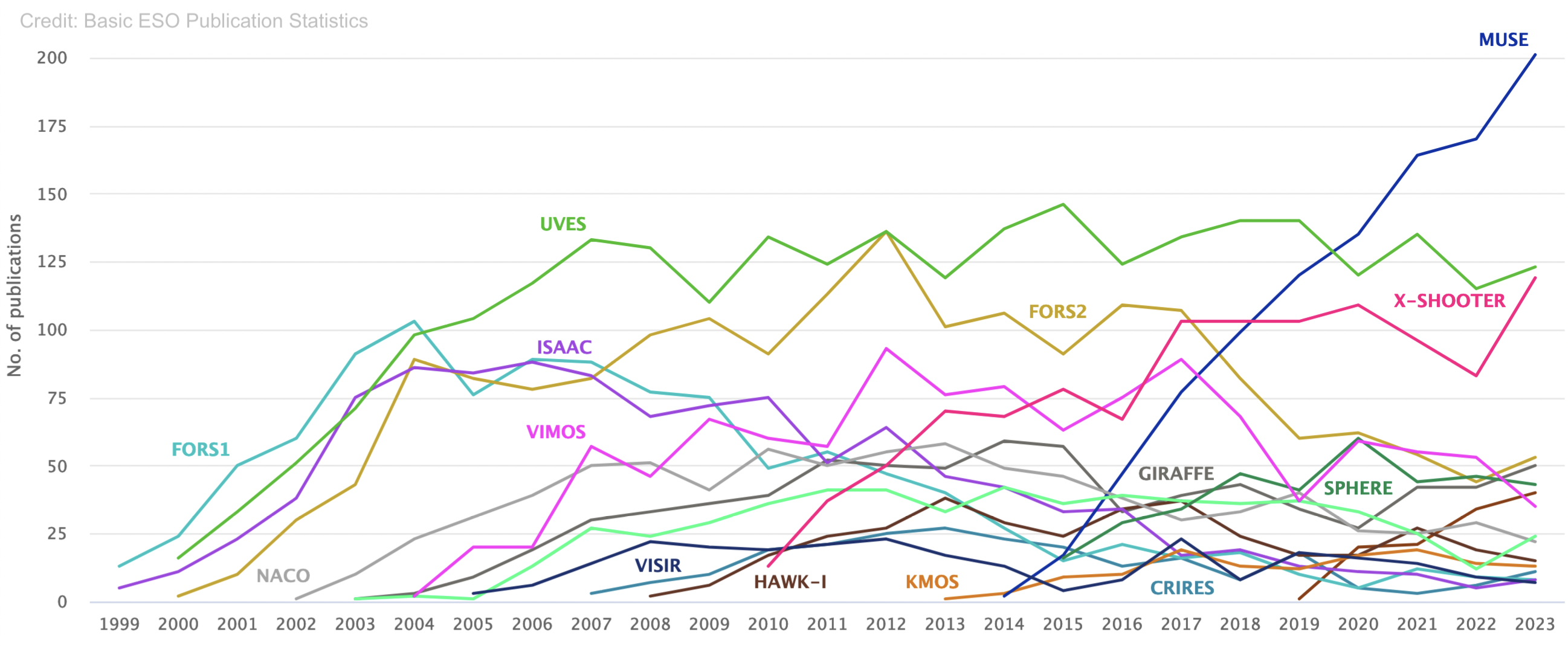}
  \includegraphics[width=1.0\textwidth]{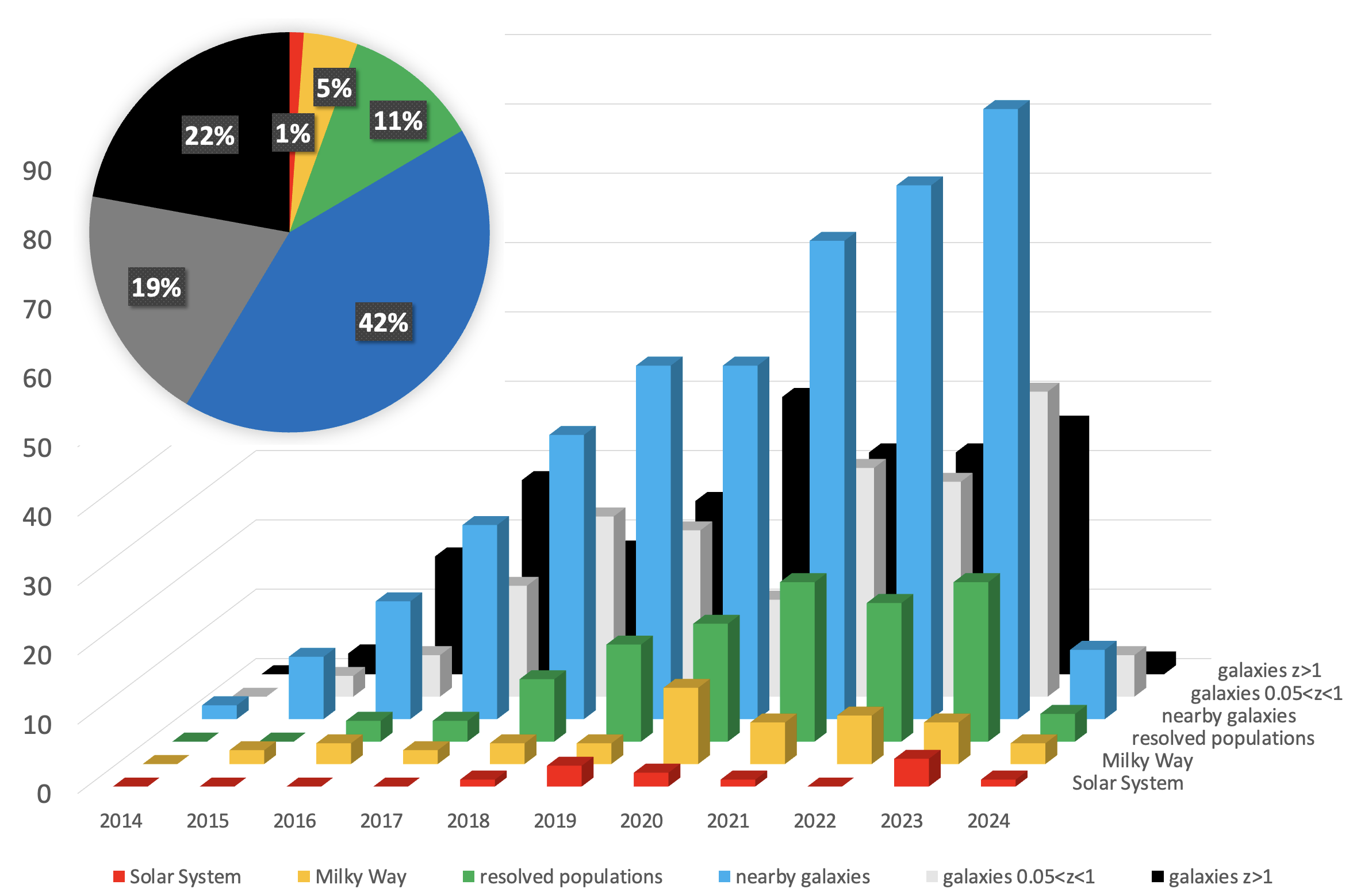}
  \caption{\small MUSE publication statistics.} 
 \label{fig:figure1}
\end{figure}

Firstly, the preponderance of extragalactic astronomy and cosmology papers is striking. This is not a complete surprise since MUSE was designed exactly for that purpose \citep{Bacon+2004}. Smaller precursor instruments such as SAURON \citep{Bacon+2001}, or PMAS \citep{Roth+2005}, and the associated galaxy surveys SAURON \citep{deZeeuw+2002} and CALIFA \citep{Sanchez+2012} have paved the way for an early uptake of extragalactic science with MUSE which is visible in the lower right panel of Fig.~\ref{fig:figure1}, where first papers appear in 2014 and 2015. 

Secondly, it is also no surprise that Milky Way and Solar System science is not strongly represented, simply because the spectral resolution R=$1800\ldots3600$ is not ideal for stellar astrophysics which, with the notable exception of hot massive stars, would typically demand R$\ge$12.000.

Thirdly, nonetheless, it is apparent that the emerging technique of IFS in crowded stellar fields, first pioneered with PMAS using the PSF-fitting tool PampelMUSE \citep{Kamann+2013}, has become a success story with growing momentum. The lag of 2 or 3 years with regard to extragalactic publications is understandable for the novelty and complexity of the technique. 

It is beyond the scope of this research note to discuss the discovery potential of MUSE, but see \citet{Bacon+2023} for an overview of scientific highlights. Lastly, instrumental properties like efficiency, calibration stability, technical reliability, support for data reduction and analysis, etc. would warrant an analysis on its own right, but again outside of the scope of this paper.

\clearpage

\bibliography{muse-pub.bib}

\end{document}